\RequirePackage{lineno}
\documentclass[prl,reprint,twocolumn,showpacs,preprintnumbers,superscriptaddress,amsmath,amssymb,floatfix]{revtex4-1}
\usepackage{graphicx}
\usepackage{amsmath}
\usepackage[usenames,dvipsnames]{color}

\usepackage[percent]{overpic}

\newcommand{\pT}{\ensuremath{p_{T}}}

\newcommand{\sqrtsn}{\ensuremath{\sqrt{s_{NN}}}}
\newcommand{\AuAu}{Au+Au}
\newcommand{\PbPb}{Pb+Pb}
\newcommand{\pp}{\ensuremath{p+p}}
\newcommand{\Aj}{\ensuremath{A_J}}
 
\graphicspath{{./figures/}}

\begin{document}

\title{Di-Jet Imbalance Measurements at $\sqrt{s_{NN}} = 200$~GeV at STAR}
\date{\today\ \textcolor{red}{version 9}}

\affiliation{AGH University of Science and Technology, FPACS, Cracow 30-059, Poland}
\affiliation{Argonne National Laboratory, Argonne, Illinois 60439}
\affiliation{Brookhaven National Laboratory, Upton, New York 11973}
\affiliation{University of California, Berkeley, California 94720}
\affiliation{University of California, Davis, California 95616}
\affiliation{University of California, Los Angeles, California 90095}
\affiliation{Central China Normal University, Wuhan, Hubei 430079}
\affiliation{University of Illinois at Chicago, Chicago, Illinois 60607}
\affiliation{Creighton University, Omaha, Nebraska 68178}
\affiliation{Czech Technical University in Prague, FNSPE, Prague, 115 19, Czech Republic}
\affiliation{Nuclear Physics Institute AS CR, 250 68 Prague, Czech Republic}
\affiliation{Frankfurt Institute for Advanced Studies FIAS, Frankfurt 60438, Germany}
\affiliation{Institute of Physics, Bhubaneswar 751005, India}
\affiliation{Indiana University, Bloomington, Indiana 47408}
\affiliation{Alikhanov Institute for Theoretical and Experimental Physics, Moscow 117218, Russia}
\affiliation{University of Jammu, Jammu 180001, India}
\affiliation{Joint Institute for Nuclear Research, Dubna, 141 980, Russia}
\affiliation{Kent State University, Kent, Ohio 44242}
\affiliation{University of Kentucky, Lexington, Kentucky, 40506-0055}
\affiliation{Lamar University, Physics Department, Beaumont, Texas 77710}
\affiliation{Institute of Modern Physics, Chinese Academy of Sciences, Lanzhou, Gansu 730000}
\affiliation{Lawrence Berkeley National Laboratory, Berkeley, California 94720}
\affiliation{Lehigh University, Bethlehem, PA, 18015}
\affiliation{Max-Planck-Institut fur Physik, Munich 80805, Germany}
\affiliation{Michigan State University, East Lansing, Michigan 48824}
\affiliation{National Research Nuclear University MEPhI, Moscow 115409, Russia}
\affiliation{National Institute of Science Education and Research, Bhubaneswar 751005, India}
\affiliation{National Cheng Kung University, Tainan 70101 }
\affiliation{Ohio State University, Columbus, Ohio 43210}
\affiliation{Institute of Nuclear Physics PAN, Cracow 31-342, Poland}
\affiliation{Panjab University, Chandigarh 160014, India}
\affiliation{Pennsylvania State University, University Park, Pennsylvania 16802}
\affiliation{Institute of High Energy Physics, Protvino 142281, Russia}
\affiliation{Purdue University, West Lafayette, Indiana 47907}
\affiliation{Pusan National University, Pusan 46241, Korea}
\affiliation{Rice University, Houston, Texas 77251}
\affiliation{University of Science and Technology of China, Hefei, Anhui 230026}
\affiliation{Shandong University, Jinan, Shandong 250100}
\affiliation{Shanghai Institute of Applied Physics, Chinese Academy of Sciences, Shanghai 201800}
\affiliation{State University Of New York, Stony Brook, NY 11794}
\affiliation{Temple University, Philadelphia, Pennsylvania 19122}
\affiliation{Texas A\&M University, College Station, Texas 77843}
\affiliation{University of Texas, Austin, Texas 78712}
\affiliation{University of Houston, Houston, Texas 77204}
\affiliation{Tsinghua University, Beijing 100084}
\affiliation{University of Tsukuba, Tsukuba, Ibaraki, Japan,}
\affiliation{United States Naval Academy, Annapolis, Maryland, 21402}
\affiliation{Valparaiso University, Valparaiso, Indiana 46383}
\affiliation{Variable Energy Cyclotron Centre, Kolkata 700064, India}
\affiliation{Warsaw University of Technology, Warsaw 00-661, Poland}
\affiliation{Wayne State University, Detroit, Michigan 48201}
\affiliation{World Laboratory for Cosmology and Particle Physics (WLCAPP), Cairo 11571, Egypt}
\affiliation{Yale University, New Haven, Connecticut 06520}

\author{L.~Adamczyk}\affiliation{AGH University of Science and Technology, FPACS, Cracow 30-059, Poland}
\author{J.~K.~Adkins}\affiliation{University of Kentucky, Lexington, Kentucky, 40506-0055}
\author{G.~Agakishiev}\affiliation{Joint Institute for Nuclear Research, Dubna, 141 980, Russia}
\author{M.~M.~Aggarwal}\affiliation{Panjab University, Chandigarh 160014, India}
\author{Z.~Ahammed}\affiliation{Variable Energy Cyclotron Centre, Kolkata 700064, India}
\author{I.~Alekseev}\affiliation{Alikhanov Institute for Theoretical and Experimental Physics, Moscow 117218, Russia}\affiliation{National Research Nuclear University MEPhI, Moscow 115409, Russia}
\author{D.~M.~Anderson}\affiliation{Texas A\&M University, College Station, Texas 77843}
\author{R.~Aoyama}\affiliation{Brookhaven National Laboratory, Upton, New York 11973}
\author{A.~Aparin}\affiliation{Joint Institute for Nuclear Research, Dubna, 141 980, Russia}
\author{D.~Arkhipkin}\affiliation{Brookhaven National Laboratory, Upton, New York 11973}
% ...
\collaboration{STAR Collaboration}\noaffiliation

\begin{abstract}

We report the first di-jet transverse momentum asymmetry measurements
from \AuAu\ and \pp\ collisions at RHIC. The two highest-energy
 back-to-back jets reconstructed from fragments with transverse
momenta above 2 GeV/$c$ display a significantly stronger momentum
imbalance in heavy-ion collisions than in the \pp\ reference.
When re-examined with correlated soft particles included, we observe 
that these di-jets then exhibit a unique new feature
 -- momentum  balance is restored to that observed in \pp\
for a jet resolution parameter of $R=0.4$, while 
re-balancing is not attained  with a smaller value of $R=0.2$.

\end{abstract}

\maketitle

High-energy collisions of large nuclei at the Relativistic Heavy Ion Collider (RHIC) at Brookhaven National Laboratory
exceed the energy density at which a strongly-coupled medium of deconfined quarks and gluons,
the quark gluon plasma (QGP), is expected to
form~\cite{star_white,*phenix_white,*brahms_white,*phobos_white}.
Partons with large transverse momentum ($\pT\gg\Lambda_{\text{QCD}}$) resulting from hard scatterings 
provide ``hard probes'' that allow for the unique opportunity to explore the QGP tomographically.
Such scatterings occur promptly ($\sim1/p_T$) in the initial stages of the collision,
and can thus probe the evolution of the medium. 
The scattered partons separate and fragment into back-to-back clusters of collimated hadrons known as jets. 
Jet \pT\ distributions in proton-proton (\pp) collisions at RHIC are well-described by perturbative quantum chromodynamics (pQCD)
and can be used as a calibrated reference for studies of medium-induced jet modifications~\cite{star_jet}. 

Production of high-\pT\ hadrons, serving as a jet proxy, 
was first found to be highly suppressed at RHIC in single-particle measurements compared to scaled \pp\ collisions.
Moreover, particle yields on the recoil side of high-\pT\ triggered di-hadron correlations
exhibited a shift from high to low energy~\cite{star_zt}.
These observations established the energy dissipation of fast-moving partons as a key signature of a dense partonic medium,
known as the \emph{jet quenching effect}~\cite{Gyulassy:1990ye,Majumder:2010qh}.
Most theoretical explanations of light quark and gluon jet quenching in heavy-ion collisions,
while differing in details, identify pQCD-type radiative energy loss (\emph{gluon bremsstrahlung})
as the dominant mechanism. Inherent to these frameworks is the qualitative feature that the jet structure
is softened and broadened with respect to vacuum expectations \cite{Gyulassy:1990ye,Majumder:2010qh,PhysRevC.90.014909,Qin:2015srf}. 
%
Advances in jet-finding techniques~\cite{fastjet}, and the proliferation of high-\pT\ jets at the higher energies accessible
at the Large Hadron Collider (LHC) with a higher center-of-mass energy per nucleon pair, have made it possible
to study fully reconstructed jets in heavy-ion collisions for the first time~\cite{Adam:2015ewa,cmsjet,Aad:2012vca}.
Inclusive jet spectra in the most central (head-on) lead-lead (\PbPb) collisions at a center-of-mass energy per nucleon pair
of \sqrtsn=2.76~TeV were found to be clearly suppressed when compared
to scaled \pp\ or scaled peripheral (glancing) \PbPb\ collisions at the same collision energy.
This suppression occurred independently of jet $p_T$ for jets with $\pT\sim40-210$~GeV/$c$,
and even for jets reconstructed with a resolution parameter as large as $R=0.5$
(while the exact meaning of $R$ is algorithm-specific, 
for the anti-$k_T$ algorithm used throughout this Letter, it typically corresponds to roughly
circular clusters of radius $R$ in $\Delta\,R=\sqrt{\Delta\,\phi^2+\Delta\,\eta^2}$
where $\Delta\,\phi$ is the relative azimuthal angle and $\Delta\,\eta$ the relative pseudorapidity).

Recently, analyses of di-jet pairs revealed a striking energy imbalance for highly energetic
back-to-back jet production~\cite{atlasjet,cmsjet}.
The reported imbalance observable is defined as
\begin{align}
  \label{eq:1}
  A_{J}\equiv (p_{T,\text{lead}}-p_{T,\text{sublead}})/(p_{T,\text{lead}}+p_{T,\text{sublead}})
\end{align}
where $p_{T,\text{lead}}$ and $p_{T,\text{sublead}}$ are the transverse momenta of the leading and sub-leading (highest and second-highest $p_T$) jet,
respectively, in the di-jets that are required to be approximately back-to-back.
In this observable, detector effects in the determination of jet $p_T$ affect
numerator and denominator in a similar manner and thus cancel out to first order.
It is therefore less sensitive to effects of the underlying event
than inclusive measurements and other di-jet observables.
Furthermore, when di-jets with large energy imbalance were examined at the LHC,
much of the \emph{lost energy} of these  jets seemed to re-emerge as low momentum particles 
emitted at large angles with respect to the di-jet axis~\cite{Chatrchyan:2012nia,Aad:2012vca,Khachatryan:2016erx}.

By contrast, at RHIC energies, measurements based on correlations of hadrons with leading reconstructed jets or non-decay (direct) photons
indicate that the lost energy remains much closer to the jet axis~\cite{jhcorr, phenix_gdir_jet},
suggesting only a moderate broadening of the jet structure for all but the softest constituents.
The difference between the RHIC and LHC energy results could be due to a number of different reasons;
both the details of the experimental analyses and the mean parton kinematics being probed at the two facilities differ significantly.
In addition, the LHC results specifically focus on di-jets with a large energy imbalance on an individual event-by-event basis,
whereas published RHIC measurements based on statistical correlations require treatment of an ensemble-based background.

In this Letter, we present the first di-jet imbalance measurement in central gold-gold (\AuAu) collisions at RHIC,
thus allowing a more direct comparison to jet quenching measurements at the LHC. 
The data used in this analysis were collected by the STAR detector in \pp\ and \AuAu\ collisions at
$\sqrt{s_{NN}}=200$~GeV in 2006 and 2007, respectively.
Charged tracks are reconstructed with the Time Projection Chamber (TPC)~\cite{Anderson:2003ur}.
The transverse energy ($E_T$) of neutral hadrons is included by measuring the energy deposited in the Barrel Electromagnetic Calorimeter (BEMC)~\cite{Beddo:2002zx},
which has a tower size of $0.05 \times 0.05$ in azimuth $\phi$ and pseudorapidity $\eta$.
To avoid double-counting, the energy deposited by charged hadrons in the BEMC is accounted for by full hadronic correction,
in which the transverse momentum of any charged track that extrapolates to a tower is subtracted from the transverse energy of that tower.
Tower energies are set to zero if they would otherwise become negative via this correction. 
While full hadronic correction is an overly conservative way to avoid double-counting energy from charged tracks,
it has been found to be the most robust approach~\cite{Adamczyk:2014ozi}.
Both the TPC and the BEMC uniformly cover the full azimuth and a pseudorapidity range of $|\eta|<1$.
Events were selected by an online high tower (HT) trigger, which required an uncorrected $E_T > 5.4$ GeV in at least one BEMC tower.
In \AuAu\ collisions, only the most central 20\% of the events are analyzed,
where event centrality is a measure of the overlap of the colliding nuclei, determined by the raw charged particle multiplicity
in the TPC within $|\eta| < 0.5$.
Events are restricted to have a primary vertex position along the beam axis of $|v_z| < 30$~cm.
Tracks are required to have more than 52\% of available points measured in the TPC (up to 45), and a minimum of 20,
a distance of closest approach (DCA) to the collision vertex of less than 1 cm,
and pseudorapidity within $|\eta| < 1$. 

Jets are reconstructed from charged tracks measured in the TPC and neutral particle information recorded by the BEMC,
using the anti-$k_{T}$ algorithm from the FastJet package~\cite{fastjetArea,fastjet} with resolution parameters $R = 0.4$ and $0.2$.
The reconstructed jet axes are required to be within $|\eta| < 1-R$ to avoid partially reconstructed jets at the edge of the acceptance.
In this analysis, the initial definition of the di-jet pair considers only tracks and towers with $p_{T} > 2$ GeV/$c$~in the jet reconstruction.
This is done to minimize the effects of background fluctuations and combinatorial jets not originating from an initial hard scatter,
and to make an average background energy subtraction unnecessary.
We will refer to this selection as (di-)jets with ``hard cores'', as most of their energy is carried by just a few high-$p_T$ constituents.
The event-by-event background energy density $\rho$ is determined
as the median of $p_T^{\text{jet,rec}} / A^{\text{jet}}$ of all but the two leading jets,
using the $k_T$ algorithm with the same resolution parameter $R$ as in the nominal jet reconstruction~\cite{fastjet}.
The area $A^{\text{jet}}$ of jets is also found with the FastJet package (using active ghost particles).
At RHIC energies, the median background energy density $\langle \rho \rangle$ when only particles
with  $p_{T} > 2$ GeV/$c$~are considered is 0.
Hence no event-by-event $\rho$ subtraction is applied for these ``hard-core'' jets.
The small residual influence of background fluctuations is captured
by embedding the \pp\ reference hard-core jets into an \AuAu\ event (after reconstruction).
When, later in the analysis, the constituent cut is lowered, $\rho$ is recalculated event-by-event
and the corrected jet $p_T = p_T^{\text{jet,rec}}  - \rho A^{\text{jet}}$ is used, 
discarding jets with $p_T<0$.

The di-jet imbalance \Aj\ is initially calculated in \AuAu\ HT events for leading and sub-leading jets
fulfilling the following requirements:
\begin{itemize}
\item $p_{T,\text{lead}}>20$ GeV/$c$ and $p_{T,\text{sublead}}>10$ GeV/$c$,
\item $|\phi_\text{lead} - \phi_\text{sublead} - \pi | < 0.4$ (back-to-back).
\end{itemize}
In this Letter, jet energies are not corrected back to the original parton energies
apart from the correction for relative reconstruction efficiency differences between \AuAu\ and \pp\ described below.
In order to make meaningful quantitative comparisons between the di-jet imbalance measured in \AuAu\ to that in \pp,
it is however necessary to compare jets which have similar initial parton energies in the two collision systems,
and to take the remaining effect of background fluctuations into account.  
It was shown in \cite{jhcorr} that \AuAu\ HT leading jets are similar to \pp\ HT leading jets embedded in a \AuAu\ background.
A di-jet imbalance reference dataset is therefore constructed in this analysis via embedding \pp\ HT events into
\AuAu\ minimum bias (i.~e., without a high tower trigger) events with a 0-20\% centrality requirement identical to the HT data
(\pp\ HT $\oplus$ \AuAu\ MB). 
The heavy ion background has the potential to bias an online high tower trigger toward a higher population of low-energy jets
that would not be accounted for by the embedding. In a previous study, this effect was conservatively accounted
for with a small systematic uncertainty~\cite{jhcorr}. 
The relatively high leading jet requirement and the robustness of the observable in this analysis
further reduce a potential influence of such a bias.
A cross-check with a higher off-line trigger requirement did not show any effect beyond statistics,
and we therefore do not assign a systematic uncertainty.

The performance of the TPC and BEMC can vary in different collision systems and over time.
The relative TPC tracking efficiency in \AuAu\ is ca. $90\% \pm 7\%$ that of \pp~\cite{jhcorr},
and this difference is accounted for in the \pp\ HT $\oplus$ \AuAu\ MB during embedding by 
randomly rejecting charged \pp\ tracks with a probability given by this efficiency difference.
The uncertainty on this correction is the largest contributor to systematic uncertainty,
and it is assessed by repeating the measurement with the respective minimum and maximum efficiency.
The tower efficiency in \AuAu\ collisions relative to \pp\ collisions is $98\% \pm 2\%$~\cite{jhcorr},
and its contribution to systematic uncertainties is negligible compared to the respective TPC uncertainty.
The systematics due to the relative tower energy scale ($100\% \pm 2\%$) is again assessed 
via the embedding procedure by increasing or decreasing the  $E_T$ of all \pp\ towers by 2\%. 
These two variations constitute the systematic uncertainty on differences between 
\AuAu\ and embedded \pp\ as discussed in this Letter.
Their quadrature sum is shown in colored shaded boxes in all figures.

\begin{figure}[t]
\begin{overpic}[width=.48\textwidth]
{{{R0.4_Fig1}.pdf}}
\put(11,30){
  \includegraphics[scale=0.055]
  {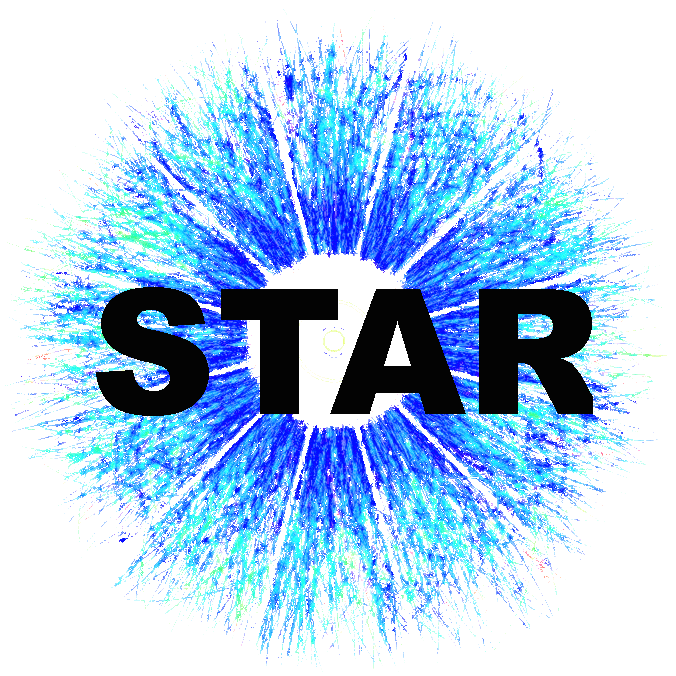}
}
\end{overpic}
  \caption{\label{fig:aj04}(Color online.)  
    Normalized \Aj\ distributions for \AuAu\ HT data (filled symbols) and \pp\ HT $\oplus$ \AuAu\ MB
    (open symbols). The red circles points are for jets found using only constituents with $p_T^{\text{Cut}}>2$~GeV/$c$
    and the black squares for matched jets found using constituents with $p_T^{\text{Cut}}>0.2$~GeV/$c$.
    In all cases $R=0.4$.}
\end{figure}

In Fig.~\ref{fig:aj04} the \Aj\ distribution from central \AuAu\ collisions for anti-$k_T$ jets with $R=0.4$ (solid red circles)
is compared to the \pp\ HT embedding reference (\pp\ HT $\oplus$ \AuAu\ MB, open circles)
for a jet constituent-$p_T$ cut of $p_{T}^{\text{Cut}} > 2$~GeV/$c$.
Di-jets in central \AuAu\ collisions are significantly more imbalanced than the corresponding \pp\ di-jets.
To further quantify this difference the p-value for the hypothesis that the two histograms represent identical distributions
was calculated with a Kolmogorov-Smirnov test on the unbinned data~\cite{Kolmogorov},
i.~e., including only the statistical uncertainties.
For an estimate of systematic effects we quote the range of minimal and maximal 
values obtained during efficiency and tower energy scale variations.
The calculated p-value $< 1\times10^{-8}$  ($4\times10^{-10}$--$1\times10^{-6}$) supports the hypothesis
that the  \AuAu\  and \pp\ HT $\oplus$ \AuAu\  data are
not drawn from the same parent  $A_J$ distributions.

In order to assess if the energy imbalance can be restored for these di-jets by including the jet constituents below 2~GeV/$c$ in transverse momentum,
the jet-finder was run again on the same events, but with a lower constituent $p_T$ cut of $p_{T}^{\text{Cut}} > 0.2$ GeV/$c$.
The di-jet imbalance \Aj\ was then recalculated for jet pairs geometrically matched to the original hard core di-jets.
For this matching, the highest \pT\ jet within $\Delta R = \sqrt{\Delta \phi^{2} + \Delta \eta^{2} }<R$
of the hard core jet was chosen. This matching has better than 99\% efficiency.
To account for the significant low-\pT\ background, this recalculation used background-corrected jet $p_T = p_T^{\text{jet,rec}}  - \rho A^{\text{jet}}$.
In the central data considered here, $\rho$ is a broad distribution with an average value of about 57~GeV/sr.
The reference \pp\ HT $\oplus$ \AuAu\ MB embedding distribution was recalculated in the same manner.
For matched jets, the role of leading and sub-leading jets is not re-enforced, so \Aj\ can now become negative;
all figures include a dashed line at 0 to guide the eye.

In Fig.~\ref{fig:aj04} the matched di-jet imbalance measured for a low constituent $p_{T}^{\text{Cut}}$ in central \AuAu\ collisions (solid black squares)
is compared to the new \pp\ HT $\oplus$ \AuAu\ MB embedding reference (open squares).
Remarkably, the \Aj\ distribution in \AuAu\ is now identical to the \pp\ data within uncertainties;
the p-value between these two distributions is 0.4 (0.2--0.6).
This observation suggests that the jet energy balance can be restored to the level of \pp\ in central \AuAu\ HT events 
for this class of di-jets if low $p_T$ constituents are included within an anti-$k_T$ jet of resolution parameter (radius) $R=0.4$.


\begin{figure}[t]
\begin{overpic}[width=.48\textwidth]
{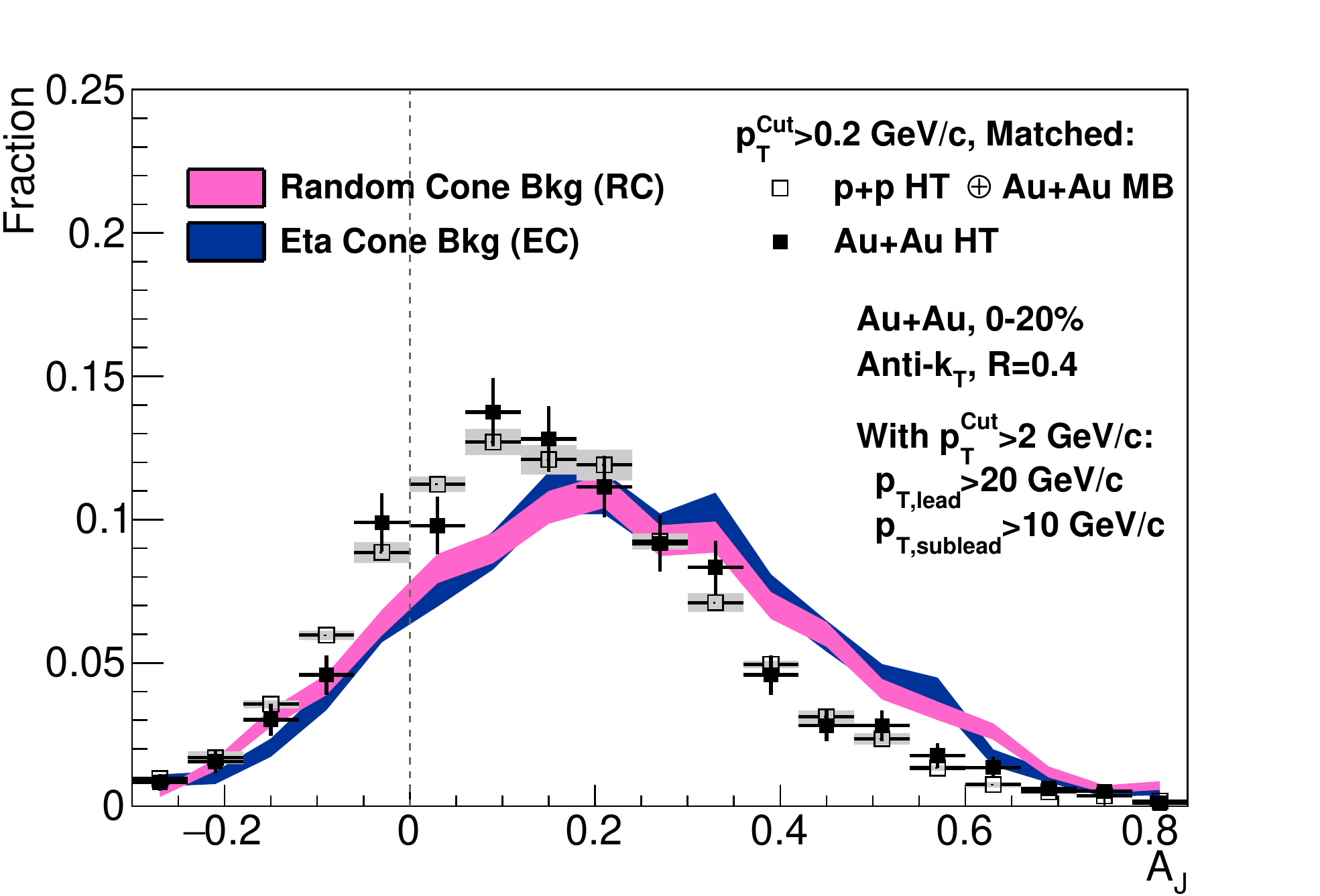}
\put(11,30){
  \includegraphics[scale=0.055]
  {STAR-logo-base-black.png}
}
\end{overpic}
  \caption{\label{fig:aj04null}(Color online.) 
    \Aj\ distributions for \AuAu\ data (filled symbols) and \pp\ HT $\oplus$ \AuAu\ MB  (open symbols)
    for low constituent $p_{T}^{\text{Cut}}$ di-jets from Fig.~\ref{fig:aj04}
    compared to \Aj\ distributions calculated assuming the RC and EC null hypotheses, 
    respectively, shown as colored bands; see the text for details.}
\end{figure}

The tremendous increase in background fluctuations below 2~GeV/$c$
could lead to an artificial di-jet energy balance unrelated to potential modifications in the jet fragmentation.
In the limit of infinitely high background fluctuations, the correlated signal could be washed out to be indistinguishable.
To estimate the magnitude of this effect, we employed two different \emph{null hypothesis} procedures.
First, we embedded the \AuAu\ HT di-jets reconstructed with $p_{T}^{\text{Cut}} > 2$~GeV/$c$
(closed red markers in Fig.~\ref{fig:aj04}) into \AuAu\ MB events with a low constituent $p_T^{\text{Cut}}>0.2$~GeV/$c$,
re-performed the jet finding and matching and re-calculated \Aj.
This procedure explicitly disallows for any balance restoration via correlated signal jet constituents since the jet is embedded into a different random event.
We refer to this method as the Random Cone (RC) technique.
In the second method, in order to account for potential non-jet correlations within the event,
we embed the same di-jet pairs as in the RC method into a different \AuAu\ HT event
with a found di-jet pair, at the same azimuth position but randomly offset in pseudorapidity by at least $2\times R$.
This ``Eta Cone'' method (EC) preserves potential background effects due to azimuthal correlations of the underlying event with the jet
while also excluding any potential jet-like correlation below 2~GeV/$c$.
Both of these methods are compared to the measured matched \Aj\  distribution with low $p_{T}^{\text{Cut}}$ in Fig.~\ref{fig:aj04null}.
We conclude that background fluctuations do smear the $A_J$ signal significantly with an overall effect toward balancing.
However, the resulting distributions still show much higher imbalance and significant shape differences compared to the measured signal.
This smearing cannot alone account for the magnitude of the rebalancing, confirming that the energy restored via low $p_T$ constituents
is correlated with the jet fragmentation.


\begin{figure}[thb]
\begin{overpic}[width=.48\textwidth]
{{{R0.2_Fig}.pdf}}
\put(11,30){
  \includegraphics[scale=0.055]
  {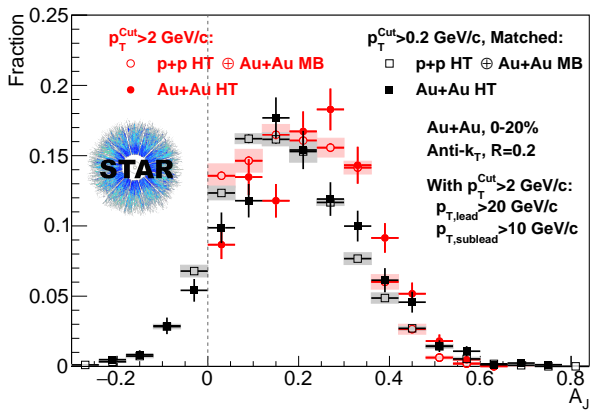}
}
\end{overpic}
  \caption{\label{fig:aj02}(Color online.)
    Repetition of the analysis shown in Fig.~\ref{fig:aj04} with a smaller resolution parameter $R=0.2$.
    Normalized \Aj\ distributions for \AuAu\ HT data (filled symbols) and \pp\ HT $\oplus$ \AuAu\ MB
    (open symbols). The red circles are for jets found using only constituents with $p_T^{\text{Cut}}>2$~GeV/$c$
    and the black squares are for matched jets found using constituents with $p_T^{\text{Cut}}>0.2$~GeV/$c$.
  }
\end{figure}

In order to assess if the observed softening of the jet fragmentation is accompanied by a broadening of the jet profile,
a measurement of the di-jet imbalance with a resolution parameter of $R=0.2$ was performed in an analogous fashion
to the measurement described above.
As shown in Fig.~\ref{fig:aj02}, narrowing the cone to $R=0.2$ leads to significant differences between 
central \AuAu\ and embedded \pp\ for jets with hard cores,
with a p-value of $1\times10^{-8}$ ($1\times10^{-9}$--$3\times10^{-7}$).
Including soft constituents down to 0.2~GeV/$c$
is no longer sufficient to restore the imbalance to the level of the \pp\ reference.
This continued disparity between the \pp\ and \AuAu\ data is supported by a calculated
p-value of $7\times10^{-8}$ ($2\times10^{-8}$--$4\times10^{-7}$).

In all descriptions of the QGP, energy redistribution via gluon bremsstrahlung is dependent on in-medium path length. 
Requiring high-\pT\ hadrons in the measured final state therefore imposes a significant bias toward 
production near the surface of the fireball, a paradigm known as Surface Bias.
Previous STAR jet-hadron measurements are well-captured by YaJEM-DE, a
Monte Carlo model of in-medium shower evolution that predicts just such a surface bias 
for the same leading jet selection as used in this Letter~\cite{jhcorr,PhysRevC.87.024905}. 


The initial hard core di-jet selection places hard hadron requirements on the recoil jet in addition to those on the leading jet.
In the surface bias picture, they are therefore expected to display a pronounced preference toward almost tangential di-jets,
probes that graze the medium with a shorter but finite in-medium path-length
compared to the unbiased di-jet selection at LHC energies~\cite{Renk:2012cx}.
Correlation measurements with two hard particles as jet proxies support the presence of such a 
tangential bias as well~\cite{Adamczyk:2012eoa}.
Our measurements of clearly modified jets whose ``lost'' energy can nevertheless be recovered
within a comparatively narrow cone are qualitatively consistent with this picture.

The qualitative change in the di-jet imbalance for smaller $R$ jets as reported in this letter is the first step towards enabling
\emph{Jet Geometry Engineering} of jet production points which will allow control over the path lengths and interaction probabilities
of jet quenching effects within the colored medium.  
In addition it would be very interesting to repeat this \Aj\ study with ``hard core'' di-jets at the LHC to see if a similar energy loss pattern
is observed when similar jet pairs are selected.
Comparison and combined analysis of these new RHIC results and current published LHC measurements will already 
enable new and enhanced constraints to be placed on the dynamics underlying modified fragmentation and energy dissipation in heavy-ion collisions.


In conclusion, we reported the first \Aj\ measurement performed at $\sqrtsn=200$~GeV.
A selection of di-jet pairs with hard cores is probed.
For a resolution parameter of $R=0.4$, a clear increase in di-jet momentum imbalance is observed compared to a \pp\ baseline
when only constituents with $p_{T}^{\text{Cut}}>$ 2 GeV/$c$ are considered.
When allowing softer constituents down to $p_{T}^{\text{Cut}} >0.2$~GeV/$c$,
the energy balance becomes the same within errors as the one measured in \pp\ data.
By contrast, repeating the same measurement with a smaller resolution parameter of $R=0.2$
leads to significant remaining momentum imbalance even for jets with soft constituents.
The results are the first indication that at RHIC energies it is possible to select a sample of di-jets
that clearly lost energy via interactions with the medium but whose lost energy re-emerges
 as soft constituents accompanied with a small, but significant, broadening of the jet structure compared to 
\pp\ fragmentation. 
The above observations are consistent with the qualitative expectations of pQCD-like radiative energy loss in the hot, dense medium created at RHIC.

We thank the RHIC Operations Group and RCF at BNL, the NERSC Center at LBNL, the KISTI Center in
Korea, and the Open Science Grid consortium for providing resources and support. This work was 
supported in part by the Office of Nuclear Physics within the U.S. DOE Office of Science,
the U.S. NSF, the Ministry of Education and Science of the Russian Federation, NSFC, CAS,
MoST and MoE of China, the National Research Foundation of Korea, NCKU (Taiwan), 
GA and MSMT of the Czech Republic, FIAS of Germany, DAE, DST, and UGC of India, the National
Science Centre of Poland, National Research Foundation, the Ministry of Science, Education and 
Sports of the Republic of Croatia, and RosAtom of Russia.

\bibliographystyle{apsrev4-1}
\bibliography{ref}

\end{document}